\documentclass[10pt,a4paper,twocolumn]{article}
\usepackage[utf8]{inputenc}
\usepackage{amsmath}
\usepackage{amsfonts}
\usepackage{amssymb}
\usepackage{graphicx}
\usepackage{yfonts}
\usepackage{xargs}
\usepackage{url}
\usepackage[pdftex,dvipsnames]{xcolor}  
\usepackage[noadjust]{cite}  
\usepackage{changes} 
\usepackage{algorithm}
\usepackage{algpseudocode}
\usepackage[font={small,it}]{caption}
\usepackage{tikz}
\usepackage{textcomp}
\usepackage{hyperref}
\usepackage{xcoffins,titling}
\usepackage[a4paper, total={6.9in, 9in}]{geometry}
\usepackage{titling}
\DeclareMathOperator*{\argmin}{arg\,min}
\DeclareMathOperator{\tr}{tr}
\usepackage{abstract}

\newcommand\copyrighttext{%
  \footnotesize \textcopyright 2016 IEEE. Personal use of this material is permitted.
  Permission from IEEE must be obtained for all other uses, in any current or future 
  media, including reprinting/republishing this material for advertising or promotional 
  purposes, creating new collective works, for resale or redistribution to servers or 
  lists, or reuse of any copyrighted component of this work in other works. \\
  DOI: \href{http://dx.doi.org/10.1109/TBME.2016.2592820}{10.1109/TBME.2016.2592820}}
\newcommand\copyrightnotice{%
\begin{tikzpicture}[remember picture,overlay]
\node[anchor=south,yshift=20pt] at (current page.south) {\fbox{\parbox{\dimexpr\textwidth-\fboxsep-\fboxrule\relax}{\copyrighttext}}};
\end{tikzpicture}%
}

\begin{document}
\title{Riemannian Geometry Applied to Detection of Respiratory States from EEG Signals: the Basis for a Brain-Ventilator Interface}
\date{}
\author{\small X.~Navarro-Sune$^{1,2}$\footnotemark[1]~, A.L. Hudson$^{3}$,,  F. De Vico Fallani$^{4-6}$,J. Martinerie$^{5,6}$, A. Witon$^{7}$,
 P. Pouget$^{5,6}$ \\ 
\small 
 M. Raux$^{1,2,8}$, T. Similowski$^{1,2,9}$ and M. Chavez$^{5,6}$\\ \vspace{-5px}
	\tiny
	$^1$Sorbonne Universit\'{e}s, UPMC Univ Paris 06, France \\ \vspace{-5px}
	\tiny
	$^2$Neurophysiologie Respiratoire Exp\'{e}rimentale et Clinique, Paris, France \\ \vspace{-5px}
        \tiny
	$^3$Neuroscience Research Australia and University of New South Wales, Sydney, Australia\\ \vspace{-5px}
	 \tiny
	$^4$INRIA Paris-Rocquencourt, ARAMIS team, Paris, France\\ \vspace{-5px}
	\tiny
	$^5$CNRS, UMR-7225, Paris, France\\ \vspace{-5px}
	 \tiny
	$^6$INSERM U1227, Institut du Cerveau et de la Moelle \'{E}pini\`ere. Paris, France\\ \vspace{-5px}
	 \tiny	
	$^7$School of Computing, University of Kent, Canterbury, UK\\ \vspace{-5px}
	 \tiny	
	$^8$Groupe Hospitalier Piti\'{e} Salp\^{e}tri\`{e}re-Charles Foix, D\'{e}partement d'Anesth\'{e}sie-R\'{e}animation, Paris, France\\ \vspace{-5px}
	 \tiny	
	$^9$Groupe Hospitalier Piti\'{e} Salp\^{e}tri\`{e}re-Charles Foix, Service de Pneumologie et R\'{e}animation M\'{e}dicale, Paris, France\\ 
}

\maketitle
\copyrightnotice
\vspace{-0.75cm}
\begin{abstract}
\textbf{Abstract--}~During mechanical ventilation, patient-ventilator disharmony is frequently observed and may result in increased breathing effort, compromising the patient's comfort and recovery. This circumstance requires clinical intervention and becomes challenging when verbal communication is difficult. In this work, we propose a brain computer interface (BCI) to automatically and non-invasively detect patient-ventilator disharmony from electroencephalographic (EEG) signals: a brain-ventilator interface (BVI). Our framework exploits the cortical activation provoked by the inspiratory compensation when the subject and the ventilator are desynchronized. Use of a one-class approach and Riemannian geometry of EEG covariance matrices allows effective classification of respiratory states. The BVI is validated on nine healthy subjects that performed different respiratory tasks that mimic a patient-ventilator disharmony. Classification performances, in terms of areas under ROC curves, are significantly improved using EEG signals compared to detection based on air flow. Reduction in the number of electrodes that can achieve discrimination can often be desirable (e.g. for portable BCI systems). By using an iterative channel selection technique, the Common Highest Order Ranking (CHOrRa), we find that a reduced set of electrodes (n=6) can slightly improve for an intra-subject configuration, and it still provides fairly good performances for a general inter-subject setting.  Results support the discriminant capacity of our approach to identify anomalous respiratory states, by learning from a training set containing only normal respiratory epochs. The proposed framework opens the door to brain-ventilator interfaces for monitoring patient's breathing comfort and adapting ventilator parameters to patient respiratory needs.
\smallskip

\end{abstract}

\footnotetext[1]{Author for correspondence: xavier.navarro@upmc.fr \\
Original article published in IEEE Trans Biomed Eng}

\section{Introduction}
Mechanical ventilation is the most frequently used life-sustaining intervention in the intensive care unit (ICU), where approximately 50\% of patients receive ventilatory support \cite{carlucci2001}. At some point in their management, many patients on mechanical ventilation (MV) are described as ``fighting their ventilator". This jargonistic expression is used to indicate a mismatch between patient respiratory efforts and ventilator ``breaths". This form of disharmony between patient and ventilator results in an increased work of breathing and is a major source of discomfort for the patient with some deleterious effects such as dyspnea and anxiety. Dyspnea and anxiety are major drivers of post-traumatic stress disorders frequently observed in patients who survive the ICU \cite{leung1997}.  Therefore, it is crucial to detect patient-ventilator disharmony as early as possible. 
Currently, this relies on monitoring of physiological signals generated indirectly (pressure, air flow) or directly (electromyography) by the respiratory muscles in response to the descending neural drive to breathe. It is generally assumed that ventilatory support should be adapted to the neural drive to breathe~\cite{Sinderby1999}. Some approaches address this issue by using the diaphragmatic EMG (e.g. the neurally adjusted ventilatory assist or NAVA~\cite{Spahija2010}). Nevertheless, these techniques fail to take into account the fact that, under certain circumstances, the automatic respiratory activity of the brain stem is supplemented by respiratory-related cortical circuits. Indeed, inspiratory loading in awake humans elicits a cortical response that can be observed in EEG signals~\cite{raux2007}. Such responses have been found to correlate with respiratory discomfort in healthy subjects fighting a ventilator~\cite{raux2007b}. These observations give rise to the prospect of an effective brain-ventilator interface (BVI) that would target the neural correlates of respiratory discomfort rather than the automatic drive to breathe~\cite{Grave2013}.  

Current neuroscience research attempts to understand how brain functions result from dynamic interactions in large-scale cortical networks, and to further identify how cognitive tasks or brain diseases contribute to reshape this organization~\cite{Varela2001}. Covariance analyses of brain data are widely used to elucidate the functional interactions between brain regions during different brain states. The relevance of covariance matrices as a feature for BCI has been already assessed~\cite{barachant2012} and they constitute a very appropriate choice given their ability to reflect spatio-temporal dynamics in EEG. In this paper, we use elements of differential geometry to evaluate the ability of EEG covariance matrices to characterize changes in respiratory states in healthy subjects.
\vspace{-0.2cm}

\subsection*{A brain-ventilator interface}
\label{bvi}
The use of brain computer interfaces (BCI) is increasingly common in clinical environments as a technology to improve patient communication and rehabilitation using brain signals \cite{mak2009}. However, the application of BCI in the respiratory context has not yet been explored.  

\begin{figure}[t!] 
   \centering
   \includegraphics[width=0.95\linewidth]{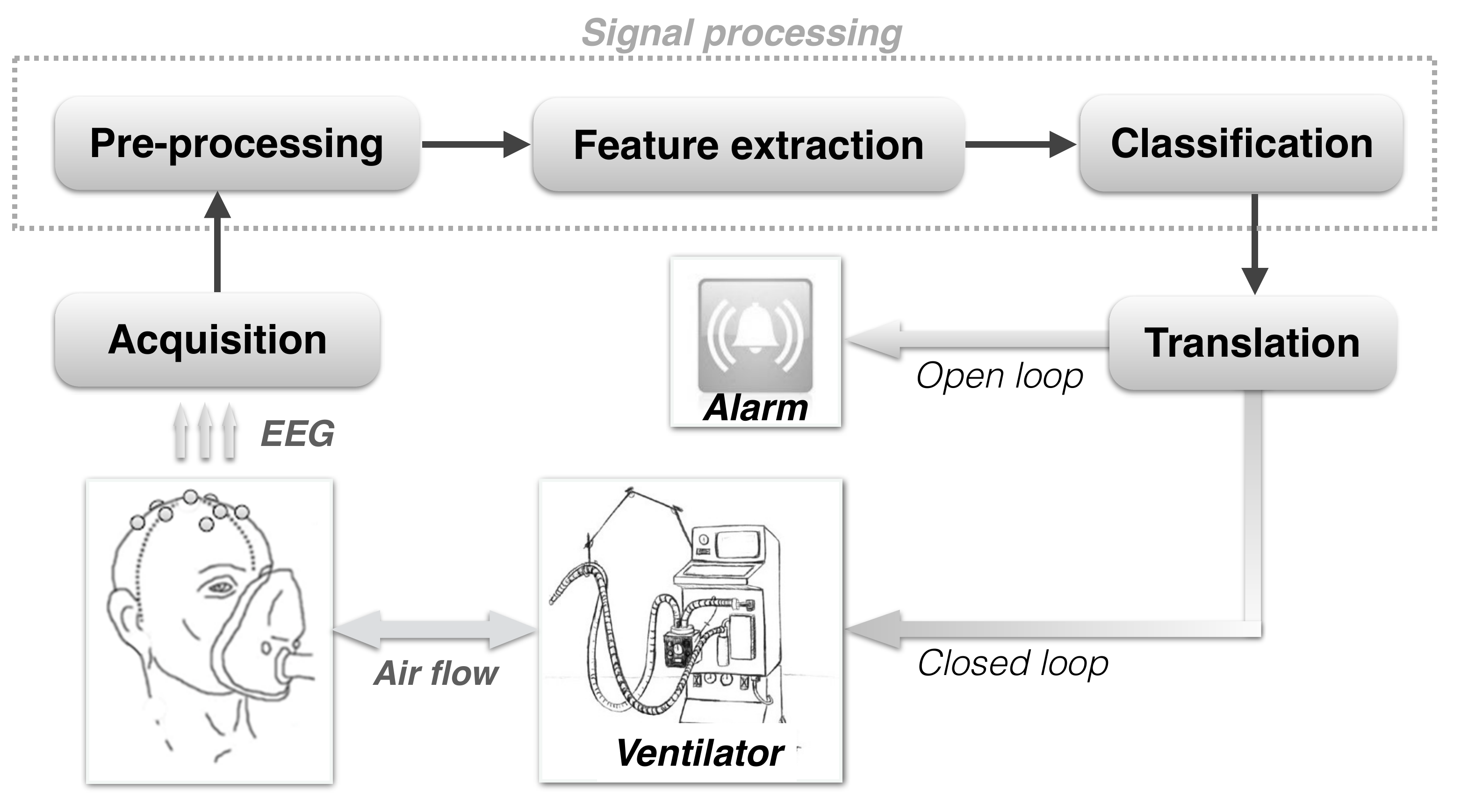} 
   \caption{Block diagram of the proposed brain-ventilator interface.}
   \label{fig:plan_bvi}
\end{figure}

In this work, we propose a framework which provides the basis of a brain-ventilator interface (see Figure \ref{fig:plan_bvi}). A first possible implementation consists of an open loop configuration that generates an output signal to trigger an alarm in the case of breathing disharmony. The second implementation is a more advanced version that could generate a continuous output signal in a closed loop to directly adapt ventilator parameters to the patient's needs. 

The different blocks in the proposed BVI are as follows:
\begin{enumerate}
\item Acquisition: Set of electrodes, amplifiers and A/D converter providing digitized EEG signals.
\item Pre-processing: Improves signal-to-noise ratios in EEG signals by applying artifact correction/rejection and/or filters.
\item Feature extraction: Sample covariance matrices (CMs) are obtained from segmented, pre-processed signals. 
\item Classification: CMs are labelled according to two possible classes: normal and altered breathing. Detection of anomalous respiratory states is achieved by one-class learning,  measurement of the distance between a number of reference matrices learned during reference condition and the CM corresponding to a particular signal epoch. Since CMs do not lie in vector space, appropriate distance metrics must consider their natural geometry.
\item Translation: External application that converts the binary signal from the classifier to an alarm or ventilator command. 
\end{enumerate}

The present paper focuses on the signal processing aspects of the BVI  --pre-processing, feature extraction and classification blocks--  as a detector of respiratory-related activities compatible with breathing discomfort. The framework is validated with EEG from healthy subjects under two breathing constraints to emulate patient-ventilator disharmony.  The reliability and performances of our method are also compared with those obtained by a Common Spatial Patterns (CSP) method in combination with Linear Discriminant Analysis (LDA).

The experimental protocol and data are detailed in Section \ref{database}. Section \ref{meth} describes the different signal processing blocks of the BVI, including our Riemann geometry based classifier and other standard classification methods that use EEG and breathing signals. Section \ref{setup} studies BVI settings for optimal detection of breathing discomfort and Section \ref{res} provides the experimental results and evaluation of the BVI. Finally, we conclude the paper with a discussion in Section \ref{concl}.

\section{Database}
\label{database}
The database is composed of nine healthy subjects (21 - 29 years; 5 women) with no prior experience with respiratory or neurophysiology experiments (for more details see \cite{hudson2016}). According to the declaration of Helsinki, written informed consent was obtained from each subject after explanation of the study, which was approved by the local institutional ethics committee (Comit\'e de Protection des Personnes Ile-de-France VI, Groupe Hospitali\`ere Piti\'e-Salp\^etri\`ere, Paris, France).  

Subjects were sitting in a comfortable chair and breathed continuously through a mouthpiece. They were asked to avoid body and head movements. They were distracted from the experimental context by watching a movie during the entire experiment, on a screen placed in front of them. To minimize emotional interference, the movie was a neutral animal documentary. It is worthy to note that, even though many ICU patients are supine, clinical practice guidelines recommend that mechanically ventilated patients in the ICU should be semi-supine, or semi-seated, to decrease the risk of respiratory infections and increase patient comfort~\cite{dodek2004}.

Electroencephalographic activity was recorded via surface electrodes (Acticap, BrainProducts GmbH, Germany) using 32 electrodes according to the standard 10-20 montage and sampled at 2500 Hz. Impedance between electrodes and skin were set below 5 k$\Omega$.  The mouthpiece was connected to a pneumotachograph (Hans Ruldoph Inc., MO, USA) and a two-way valve to measure air flow and attached, when required, an inspiratory load (range 18-25 cmH20).  

The experiment was designed to activate cortical regions by altered breathing and consisted of three parts:
\begin{enumerate}
\item Normal, spontaneous ventilation (SV condition). Breathing is controlled automatically by the autonomous nervous system without cortical contribution.
\item Voluntary brisk inhalations or sniffs (SN condition). Breathing movements are planned before execution, thus motor and pre-motor cortical regions are solicited. 
\item Inspiratory loaded breathing (LD condition). Ventilatory muscles perform a supplementary effort to overcome an inspiratory threshold load to maintain adequate air flow, a condition known to engage cortical networks~\cite{raux2007}.
\end{enumerate}

In contrast to the SV condition, where breathing is comfortable, LD is associated with respiratory discomfort. In a clinical context, SV would correspond to patient-ventilator harmony, whereas LD condition would correspond to patient-ventilator disharmony. SN can be considered as a positive control condition where cortical control is expected. For all subjects, 10 minutes of EEG was recorded for each condition, i.e. 10 min of SV, 10 min of SN and 10 min of LN. The experiments were well tolerated by all subjects and no intervention was necessary to modify the amount of load or sniff pattern during the recordings. Hence, during 10-minute recordings for each condition, no relevant changes occurred within each of these 10-min blocks regarding experimental conditions or subject behavior.

\section{Methods}
\label{meth}

\subsection{Pre-processing}
\label{subsec:preproc}
The purpose of this block is to enhance motor cortical activity, whose main rhythms are within $\mu$ and $\beta$ bands \cite{pfurtscheller1999}. Therefore, signals were band-pass filtered by a linear phase FIR filter (see Section \ref{ch_freq} for more details regarding frequency selection).
Data segments with artifacts due to repetitive eye blinks and ocular movements were visually detected and removed from the original EEG dataset. Then, signals were down-sampled to 250 Hz and segmented in 5 second sliding, 50\% overlapped windows to reduce computational cost in subsequent blocks. This time interval is in concordance with the slow breathing dynamics (a breath every 2.5 to 5 seconds).

\subsection{Feature extraction}
As mentioned above, the basis to classify the breathing state in the BVI are sample covariance matrices. The feature extraction block processed EEG data in epochs of $N_t$ samples and $N_c$ channels, as a matrix $\mathbf{X}  \in \mathbb{R}^{N_t \times N_c}$, and then transformed to a sample covariance matrix $\mathbf{P} \in \mathbb{R}^{N_c \times N_c}$. The latter was computed by the unbiased estimator:
\begin{equation}
\mathbf{P} = \frac{1}{N_t-1} \mathbf{X} \mathbf{X}^{T} \: ,
\end{equation}
where the superscript $T$ denotes the matrix transpose. 

By construction, $\mathbf{P}$ are symmetric and positive-definite matrices (SPD) that do not lie in a vector space but  on a Riemannian manifold \cite{moakher2005}. Therefore, previous methods defined on a Euclidean structure are not longer adequate. The correct manipulation of these SPD matrices requires the application of Riemannian geometry concepts, described in Section \ref{riemann}. 

\subsection{Classification}
\label{classif}
\subsubsection{Classification of covariance matrices}
Thanks to the ability of covariance matrices to capture EEG spatial dynamics, a one-class approach can be chosen to solve the classification problem in a BVI. 

In one-class algorithms, the availability of the labels from only one class is enough to classify instances from a second class, as the latter can be mapped in a different (distant) region of the space representation. In our framework,  motor cortex activation provoked during uncomfortable breathing should be different to that underlying normal, comfortable breathing (absence of motor cortical activity).

The objective of the classifier is to learn from $L$ data samples of the reference class SV  ($\mathbf{X}_l^{(0)}, l=1 \ldots L$) in order to label new trials $\mathbf{X}$ in two classes, $\mathcal{C}(\mathbf{X}) = \lbrace 0,1 \rbrace $, which correspond to breathing comfort (SV condition) and discomfort (SN or LD conditions), respectively.
  
During the learning process, the algorithm first finds $K$ matrices, subsequently called prototypes, serving as reference to perform classification. Those prototypes constitute $K$ centers from the ensemble of covariance matrices $\mathbf{P}_l^{(0)}$ and they are estimated by means of a general $K$-means clustering algorithm: 
 \begin{enumerate}
\item Initialize the prototypes $\mathbf{C}_k$,  $k=1 \ldots K$ by random selection of $K$ matrices from $\mathbf{P}_l^{(0)}$,
\item For each sample covariance matrix, compute its distance $d$ to all the prototypes and assign it to the closest one in order to form the cluster $\mathbf{S}_k$:
\begin{displaymath}
\mathbf{S}_k = \left\{ \mathbf{P}_n: d(\mathbf{P}_n, \mathbf{C}_k)\leqslant d(\mathbf{P}_n, \mathbf{C}_j)\forall j, 1\leqslant j \leqslant K \right\}
\end{displaymath}
\item Update each prototype by averaging points in $\mathbf{S}_k$ as 
\begin{displaymath}
\mathbf{C}_k \leftarrow \argmin_{\mathbf{C}}\sum_{\mathbf{C}_i \in \mathbf{S}_k} d (\mathbf{C}, \mathbf{C}_i)
\end{displaymath}
For points lying in a vector space, $\mathbf{C}_k$ corresponds to the arithmetic mean of points in $\mathbf{S}_k$
\item Go to step (2) until convergence (e.g. the assignments no longer change) is achieved.
 \end{enumerate}

The resulting prototypes $\mathbf{C}_k$ represent the Class $0$, and are then used to classify the new unlabelled data $\mathbf{X}_i$ (sub-index $i$ denotes the number of the $N_t$-sample window) according to:
\begin{equation}
\label{eq:class}
\mathcal{C}(\mathbf{X}_i) =
  \begin{cases}
    0 & \text{if }  \delta_i  \leq  \kappa   \\
    1 &  \text{otherwise}
  \end{cases},
\end{equation}
where $\delta_i$ is the distance to the closest reference prototype, $\delta_i = \min_{k}  d (\mathbf{P}_i, \mathbf{C}_k)$; and $\kappa>0$ is a scalar that can be adjusted to a performance criterion, like a statistical significance or a desired specificity/sensitivity value, for instance.

\subsubsection*{Performance metrics}
 Classification performance was measured by the area under the curve (AUC) of the receiver operating characteristic (ROC). AUC values range from 0.5 (a random classification)  to 1 (perfect classification). All AUC values were computed by applying 10-fold cross validation, excluding the learning period in the classification.  

\begin{figure*}[t!] 
   \centering
   \includegraphics[trim = 3.5cm 0.5cm 2cm 0cm,clip=true,width=\linewidth]{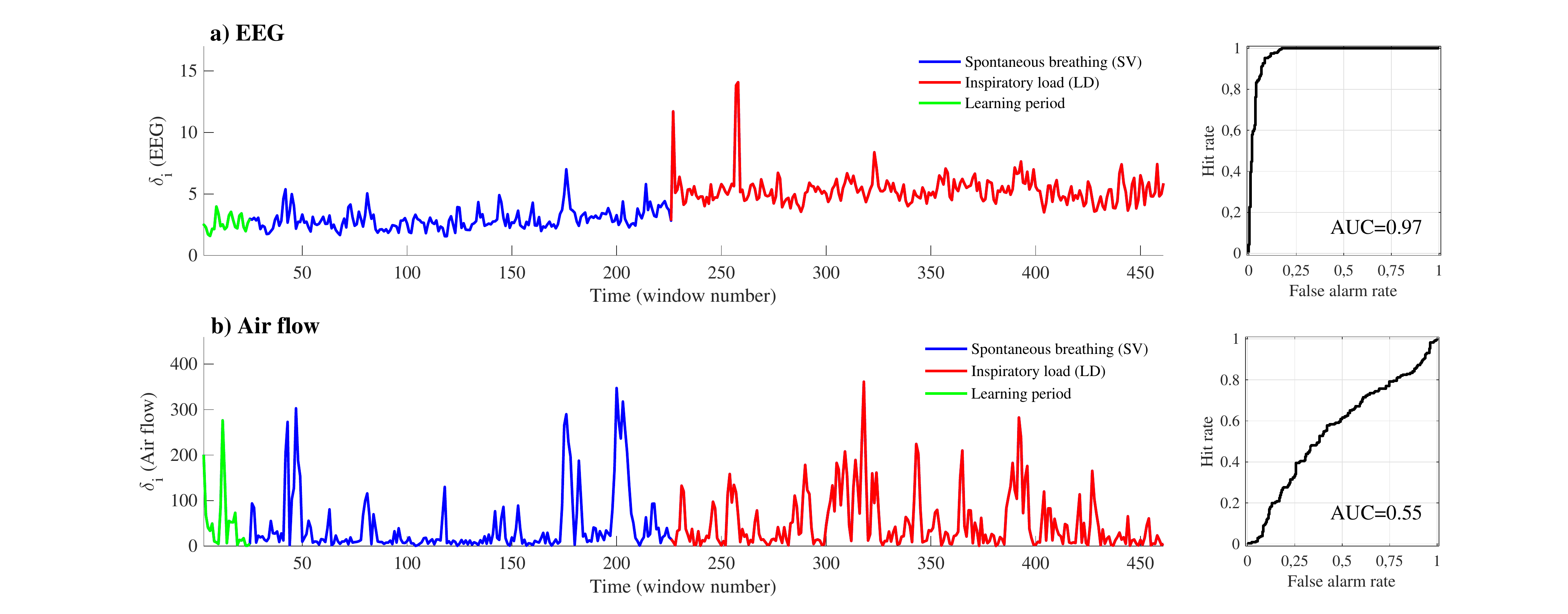}    
   \caption{Example of $\delta_i$ distances to the closest reference prototypes and the associated ROC curve obtained from a) EEG covariance matrices  (with Log-Euclidean metric, $L=25$ and $K=3$) and b) air flow, under SV (blue curves) and LD (red curves) conditions. For illustration purposes, air flow volume was used here as a feature to classify the different ventilatory conditions as done with EEG, i.e. the classifier associates the feature estimated in the $i$-th sample window to the nearest reference prototype defined in the learning period, using the Euclidean distance for the one-dimensional case. 
   }
   \label{fig:alpha_delta}
\end{figure*}

\subsection{Computation with symmetric positive definite matrices}
\label{riemann}
Within an Euclidean framework, the $K$-means algorithm divides the dataset into groups and attempts to minimize the Euclidean distance between samples labeled to be in a cluster and a point designated as the arithmetic mean of that cluster. Nevertheless, the SPD manifold is not a linear space with the conventional matrix addition operation. A natural way to measure closeness on a manifold is by considering the geodesic distance between two points on the manifold~\cite{bergerBOOK}. Such distance is defined as the length of the shortest curve connecting the two points.  As an example, consider the Earth's surface as a manifold: the Riemannian distance between the two poles is given by a meridian, while the Euclidean distance corresponds to the straight line going through the Earth's core from pole to pole.

In the space of SDP matrices, the clustering algorithm must minimize the geodesic distances between each point of the manifold (the covariance matrices $\mathbf{P}_n$) and the reference matrix $\mathbf{C}_k$. Each cluster center can be then obtained by an averaging process that employs the intrinsic geometrical structure of the underlying set.

Derived from different geometrical, statistical or information-theoretic considerations, various distance measures have been proposed for the analysis of SPD matrices~\cite{pennec2006, arsigny2006, dryden2009, vemuri2011, cherian2013}. Although many of these distances try to capture the non-linearity of SPD matrices, not all of them are geodesic distances.

The space of SPD matrices, $\mathcal{P}_{N_c}$, constitutes a differentiable Riemannian manifold $\mathcal{M}$ of dimension $Nc(Nc+1)/2$. At any point $\mathbf{Q} \in \mathcal{P}_{N_c}$ there is a tangent Euclidean space, $T_{P_{N_c}}$. Let two points $\mathbf{T}_1$ and $\mathbf{T}_2$ be two points on the tangent space (e.g. the projection of two SPD matrices), the scalar product in the tangent space at $\mathbf{Q}$ is defined by  
\begin{equation}
\label{e1}
\langle \mathbf{T}_1,\mathbf{T}_2 \rangle_\mathbf{Q} = \tr (\mathbf{T}_1\mathbf{Q}^{-1}\mathbf{T}_2\mathbf{Q}^{-1}),
\end{equation}
 that depends on point $\mathbf{Q}$.

The logarithmic map that locally projects a covariance matrix $\mathbf{P}$ onto the tangent plane is given by:
\begin{equation}
\label{e2}
\mathrm{Log}_\mathbf{Q} (\mathbf{P}) = \mathbf{S_Q} = \mathbf{Q}^{\frac{1}{2}} \mathrm{logm} (\mathbf{Q}^{-\frac{1}{2}} \mathbf{P} \; \mathbf{Q}^{-\frac{1}{2}}) \mathbf{Q}^{\frac{1}{2}}.
\end{equation}
where $\mathrm{logm}$ is the matrix logarithm operator. Projecting SDP matrices onto $T_{P_{N_c}}$ is advantageous because this tangent space is Euclidean and distance computations in the manifold can be well approximated by Euclidean distances in $T_{P_{N_c}}$. 

The inverse operation that projects a point $\mathbf{S_Q}$ of the tangent space back to the manifold $\mathcal{M}$ is given by the exponential mapping:
\begin{equation}
\mathrm{Exp}_\mathbf{Q} (\mathbf{S_Q}) = \mathbf{P} = \mathbf{Q}^{\frac{1}{2}} \mathrm{expm} (\mathbf{Q}^{-\frac{1}{2}} \mathbf{S_Q} \; \mathbf{Q}^{-\frac{1}{2}}) \mathbf{Q}^{\frac{1}{2}}
\end{equation}

In this paper, we have employed the two most widely used distance measures in the context of Riemannian manifolds: the affine-invariant distance~\cite{pennec2006} and the log-Frobenius distance, also referred to as log-Euclidean distance ~\cite{arsigny2006}. For comparative purposes, we have also used an Euclidean metric.

\subsubsection{The Euclidean metric}
On the space of real square matrices, we have the Frobenius inner product $\langle \mathbf{P}_1,\mathbf{P}_2 \rangle=\tr(\mathbf{P}_1^T\mathbf{P}_2)$ and the associated metric:
\begin{equation}
\label{dE}
d_E(\mathbf{P}_1,\mathbf{P}_2) = \| \mathbf{P}_1 - \mathbf{P}_2\|_F 
\end{equation}
Given a set of $m$ real square matrices $\mathbf{P}_1, \ldots,\mathbf{P}_m \in \mathcal{P}_{N_c}$, their arithmetic mean is given by:
\begin{equation}
\label{MdE}
\mathfrak{M}_{E}(\mathbf{P}_1, \ldots ,\mathbf{P}_m) = \frac{1}{m}\sum_{k=1}^{m} \mathbf{P}_k.
\end{equation}

\subsubsection{The affine invariant Riemannian metric}
 Using~(\ref{e1}), the Riemannian distance between two SPD matrices can be computed as:
\begin{equation}
\label{dR}
d_R(\mathbf{P}_1,\mathbf{P}_2) = \| \mathrm{logm}(\mathbf{P}_2^{-\frac{1}{2}} \mathbf{P}_1 \mathbf{P}_2^{-\frac{1}{2}}) \|_F 
\end{equation}
This metric has several useful theoretical properties: it is symmetric and satisfies the triangle inequality. Furthermore, it is scale, rotation and inversion invariant~\cite{arsigny2006, pennec2006}.

To find the mean of a set of $m$ covariance matrices $\mathbf{P}_1, \ldots,\mathbf{P}_m \in \mathcal{P}_{N_c}$, the distance $d_R$ needs to be applied in the following expression \cite{moakher2005, arsigny2007}:
\begin{equation}
\label{MdR}
\mathfrak{M}_R(\mathbf{P}_1, \ldots ,\mathbf{P}_m) = \argmin_{\mathbf{P} \in \mathcal{P}_{N_c}} \sum_{k=1}^{m} d_R(\mathbf{P}_k,\mathbf{P})^2.
\end{equation}
Although no closed-form expression exists, this (geometric) mean can be computed efficiently
using an iterative algorithm. The geometric mean in $\mathcal{P}_{N_c}$ converges into a unique solution \cite{karcher1977} and can be computed efficiently by the algorithm described in~\cite{fletcher2004}.

\subsubsection{The log-Euclidean Riemannian metric}
The log-Euclidean  distance between two SPD matrices is given by~\cite{arsigny2006}: 
\begin{equation}
\label{dLE}
d_{LE}(\mathbf{P}_1,\mathbf{P}_2) = \| \mathrm{Log}(\mathbf{P}_1)-\mathrm{Log}( \mathbf{P}_2) \|_F
\end{equation}
This metric maps SPD matrices to a flat Riemannian space (of zero curvature) so that classical Euclidean computations can be applied~\cite{arsigny2006}. Under this metric, the geodesic distance on the manifold corresponds to a Euclidean distance in the tangent space at the identity matrix. This metric is easy to compute and preserves some important theoretical properties, such as scale, rotation and inversion invariance~\cite{arsigny2006, cherian2013}.

Given a set of $m$ covariance matrices $\mathbf{P}_1, \ldots,\mathbf{P}_m \in \mathcal{P}_{N_c}$, their log-Euclidean mean exists and is uniquely determined by~\cite{arsigny2006}:
\begin{equation}
\label{MdLE}
\mathfrak{M}_{LE}(\mathbf{P}_1, \ldots ,\mathbf{P}_m) = \mathrm{Exp} \left( \frac{1}{m}\sum_{k=1}^{m} \mathrm{Log} (\mathbf{P}_k) \right)
\end{equation}

\subsection{Detection using alternate classifiers}
In this paper, we have also evaluated the detection of altered respiratory states using EEG and air flow-based features as inputs for one-class support vector machine (OCSVM) classifiers. Although one-class classification, or novelty detection, is more appropriate for the clinical monitoring of physiological condition in ventilated patients (i.e. altered breathing epochs may be detected as departures from ``normal'' breathing state), we have compared our results with those obtained from a Common Spatial Pattern (CSP) method in combination with Linear Discriminant Analysis (LDA) to detect altered breathing from EEG. CSP is a two-class oriented approach that needs a-priori labelled data from two classes previously defined (the normal and the altered breathing conditions) to train the classifier, while our one-class approach learns from a single training set containing only normal respiratory epochs. However, we have included these methods in the comparative study as they constitute a standard in brain computer interfacing \cite{lotte2007}. 

\subsubsection{One-class Support Vector Machines}
One-class Support Vector Machines is a very popular machine learning technique used as outlier or novelty detector for a variety of applications~\cite{scholkopfBook}. In OCSVM, the support vector model is trained on data that has only one class, which represents the \emph{normal} class. This model attempts to learn a decision boundary that achieves the maximum separation between the points and the origin. A OCSVM firstly uses a transformation function defined by a kernel function to project the data into a higher dimensional space. The algorithm then learns a decision boundary that encloses the majority of the projected data, and that can be applied to the outliers detection problem. More details on the algorithmic aspects of one-class SVMs can be found in~\cite{scholkopfBook}. To guarantee the existence of a decision boundary we have used here a Gaussian kernel:
\begin{equation}
 k(x_i,x_j)=\exp^{\|x_i-x_j\|^2/\sigma}.
\end{equation}
 where $x_i$ and $x_j$ denote two feature vectors and the parameter $\sigma$ is set to be the median pairwise distances among training points~\cite{quang2013}.

\subsubsection*{EEG based classifier}
For this classifier, feature vectors $x$ were extracted by computation of covariance matrices $\mathbf{P}_i$ in epoched data $\mathbf{X}_i$ followed by determination of the vector form of the upper triangular part of $\mathbf{P}_i$.

\subsubsection*{Air flow based classifier}
Air flow features were computed, as for EEG classifiers, in 50\% overlapped $N_t$-sample windows. In this case, feature vectors $x$ are composed of six air flow descriptors computed in each window \cite{abaza2009}: Peak value (l/s), average flow (l/s), total volume (l), flow variance, skewness and kurtosis. After an exhaustive feature selection procedure (testing the classifier with all $2^6$ possible combinations) using 10-fold cross validation, the greatest averaged AUCs was provided by the combination of 3 features: air flow peak value, variance and skewness. 

\subsubsection{CSP-based classification}
CSP is a widely used spatial filtering technique in BCIs to find linear combinations of electrodes such that filtered EEG maximizes the variance difference between two classes \cite{blankertz2008, haufe2014}. The computation of CSP yields a projection matrix $\mathbf{W}$ whose column vectors $w_j$ ($j=1 \ldots N_c$), called spatial filters, can be considered as an operator that transforms EEG to ``source'' components.  The matrix $\mathbf{A} = (\mathbf{W}^{-1})^T$  contains the so-called spatial patterns $a_j$ in its columns which can be considered as EEG source distribution vectors. A standard procedure to extract features in EEG epoched data ($\mathbf{X}_i$) by CSP consists of computation of the log-variances of projected $\mathbf{X}_i$ on the spatial filters~\cite{blankertz2008}. Following the recommendations in~\cite{blankertz2008}, we used the 4 most relevant spatial filters to obtain the feature vectors $x$, which were then used as inputs in a LDA classifier.

\section{BVI setup} \label{setup}

\subsection{Choice of the frequency band}
\label{ch_freq}
Following BCI designs, the BVI should exploit the oscillatory modulations on $\mu$ and $\beta$ bands (8-30 Hz) following motor and somato-sensorial cortical activations \cite{pfurtscheller1999}. Very low frequency potentials associated with voluntary movements \cite{voipio2003} may also be present during respiratory tasks as evidenced by previous work \cite{raux2007, hudson2016}.
To study the impact of specific frequency bands on the efficacy of the BVI to detect altered breathing, we compared several frequency ranges (from 0 to 30 Hz) and bandwidths (4 to 22 Hz) to tune the bandpass filter used in the preprocessing block (see Section \ref{res:freq}).

\subsection{Channel selection}
\label{ch_selelct}
In BCI, the choice of the adequate number of sensors and their location is fundamental.  
Reducing the number of irrelevant electrodes avoids over-fitting and optimizes computational costs, but also improves patient comfort and reduces installation time. Several channel selection methods can be employed to simplify the implementation of BCIs, being the most popular those based on Common Spatial Pattern (CSP)~\cite{ramoser2000, farquhar2006, wang2005} and SVMs~\cite{lal2004}. 

In addition to CSP based channel selection, we propose an iterative procedure to rank electrodes according to their discriminating power, the Common Highest Order Ranking (CHOrRa). Results obtained by both methods are compared in Section \ref{elect_rankings}).

\subsubsection{The Common Highest Order Ranking method}
CHOrRa is performed in two steps. Firstly, an electrode ranking for each subject (intra-subject rankings) is found. Then, a general ranking of the most significant electrodes is computed over all subjects (inter-subject rankings). 

To compute intra-subject ranks, a recursive backward elimination of electrodes was applied until the remaining number was 2, the minimal dimension to compute a covariance matrix (see Algorithm 1).

\begin{algorithm}
\caption{Intra-subject rank computation}\label{algo1}
\begin{algorithmic}[1]
\State initialise the subset $S$ with the initial $N_I$ electrodes
\For{each $ i=0 \ldots$ $N_I - 2$ }
    \For{each $S_j, j=1 \ldots  N_I - i$ }
      \State  $\text{AUC}_j$ = classification performance without $S_j$
    \EndFor
	 \State $O_i$ = $S_j$ that provides the minimal  $\text{AUC}_i$
	 \State remove the electrode $S_j$ from the list $S$
    \EndFor
\State \textbf{return} $O$
\end{algorithmic}
\end{algorithm}

At the end of this procedure, the list $O$ contains the indices to the $N_I$ electrodes sorted from more to less relevant in terms of classification performances for a particular subject. Combining ranking lists from all subjects, a general configuration for a ready-to-use BVI device can be chosen. We have tested two possible rank combinations:  
\begin{enumerate}
\item \textbf{Ranking aggregation}: This method determines electrode position in the ranked list, compared to a null model where all the lists are randomly shuffled~\cite{kolde2012}. Based on order statistics, this algorithm assigns a $p$-value to each electrode in the aggregated list describing the rank improvement compared to expected (a score of $1$ is assigned to channels that are preferentially ranked at the top of all lists). The procedure only takes into account the best rank, thus providing a robust re-ordering of elements~\cite{kolde2012}. 
\item \textbf{Averaged ranking}: This is a classical positional method where the total score of an electrode is simply given by the arithmetic mean of the positions in which the channel appears in each ranking list. 
\end{enumerate}

To increase rank robustness, several learning periods were used. Hence, after obtaining a set of $V$ lists ($V$-fold cross validation) per subject, the intra-subject ranking was computed according to one of the above rank combination applied to the $V$ lists of electrode rankings. 

\subsubsection*{Inter-subject ranks}
As the electrode positions leading to the maximal AUC in a particular subject isn't necessarily the same for another, we studied a common electrode configuration that could be shared by any user. This general electrode set-up is more pertinent in a clinical context as an EEG headset could be interfaced promptly. Therefore, we applied the above rank combination methods to all intra-subject rankings to obtain common and unique electrode lists.

\subsubsection{Electrode selection based on CSP}
Since CSP patterns $a_j$ ($j=1 \ldots N_c$) can be considered as EEG source distribution vectors, their associated weights can be used to find the most relevant electrodes with regard to the discrimination of the two classes~\cite{wang2005}. 

To compute intra-subject CSP-based ranks, we retained the first pattern ($a_1$) as it explains the largest variance for Class 0 (and the lowest for Class 1) and the last pattern ($a_{N_c}$), that explains the largest variance for Class 1 (and the lowest for Class 0). Within each pattern, the largest weights (absolute values) are associated with the most relevant electrodes. Hence, the list of electrode ranks was generated by selecting, alternately, in decreasing order, the first weight in $a_1$, then the first weight in $a_{N_c}$, the second weight in $a_1$ and so on as proposed by Wang et al \cite{wang2005}.

The final list of inter-subject CSP ranks were obtained by normalizing (by their maximal value) $a_1$ and $a_{N_c}$ in each subject, then finding the average of these patterns across subjects as in~\cite{ray2015}, and finally by selecting the weights as done above for intra-subject ranks.   

For consistency with CHOrRA ranks, a similar $V$-fold procedure was applied:  CSP was computed using $V$ different mean covariance matrices in Class 0. The average of the resultant $V$ projection matrices yielded the final $W$.

\subsection{Choice of prototypes and learning time}
\label{ch_prt}
As mentioned above, the prototypes are local centers that represent the space defined by $\mathbf{P}_l^{(0)}$. To this end, we adopted the $K$ prototypes provided by the general $K$-means algorithm as the centers $\mathbf{C}_k$ introduced in Section \ref{classif}. The structure of the manifold defined by $\mathbf{P}_l^{(0)}$ is unknown. Covariance matrices may be organized in a compact, uniform manner so a few centers are enough to correctly represent Class 0. On the contrary, $\mathbf{P}_l^{(0)}$ may have a complex distribution and need more centers to be correctly represented. Therefore, in order to achieve good classification, there may be an optimal number of prototypes.

To satisfactorily exploit the BVI, previous knowledge about the learning time needed to train the classifier is necessary. Limited learning time may result in a small number of covariance matrices $L$ and hence, in a poor representation of the space defined by normal breathing. Moreover, the clustering algorithm would provide a less accurate center when few CMs are available. On the other hand, the learning time should meet clinical requirements, as long set-up times in these environments are impractical.

Since the number of covariance matrices needed to represent Class 0 may impact on the optimal number of prototypes, the classification performance has to be computed by modifying $L$ and $K$. In next section, these parameters are tested experimentally to find the optimal values.

\section{Results and discussion}
\label{res}
This section shows and discusses the results of applying the BVI to detect altered respiratory condition (LD or SN) after learning a reference period of normal breathing (SV). To find the best settings in the classification block, several methods estimated the distances ($\delta_i$) to the reference condition SV. Distances were then employed to estimate the areas under ROC curves (AUC) as a measure of the classifier's performance. Plots in Figure~\ref{fig:alpha_delta} illustrate the feature distances and corresponding AUCs for detection of LD condition in one single subject.

For the EEG-based classifier, the optimal frequency band for preprocessing signals was first selected. Then, Euclidean, Riemannian and Log-euclidean distances were compared to select the best metric used for evaluating the optimal electrode configuration. Finally, the impact of learning time and the number of prototypes were also assessed. 

\begin{figure}[t!] 
   \centering
   \includegraphics[trim = 1.8cm 0cm 2cm 0cm,clip=true,width=\linewidth]{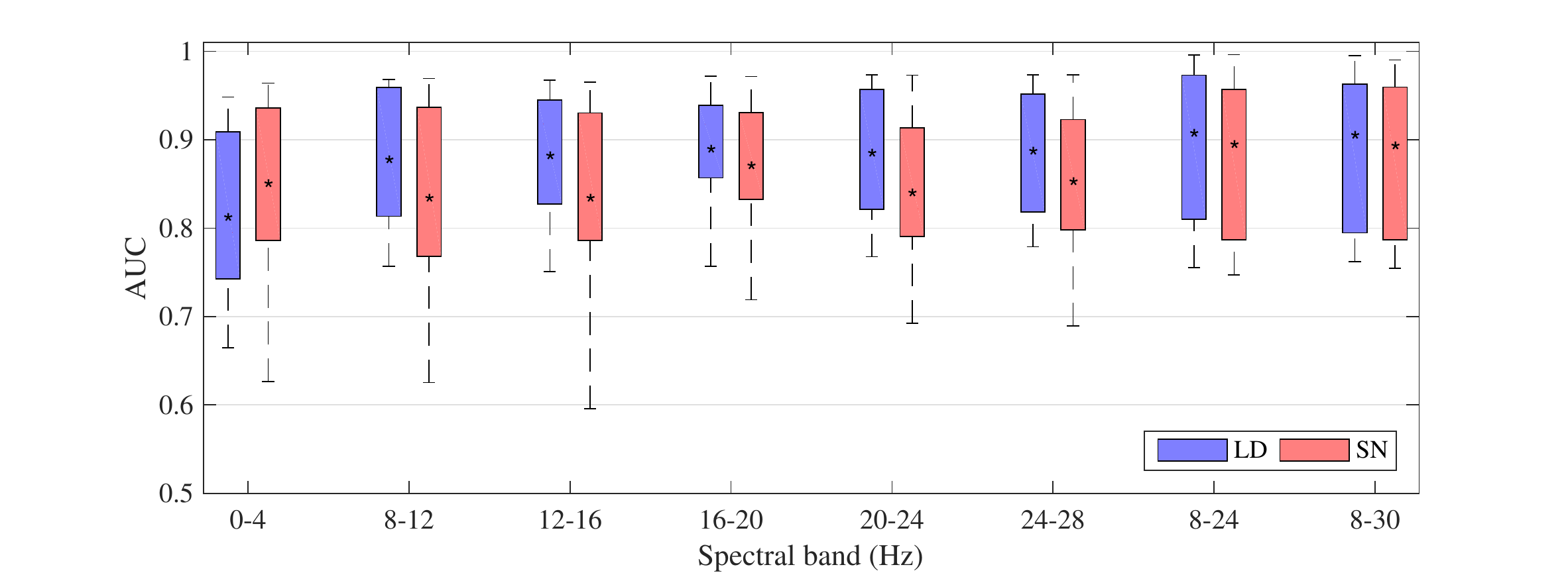} 
   \caption{Spectral profile of the BVI performances (AUCs) using Log-Euclidean distances, $K=3$ prototypes and $L=25$ covariance matrices.}
   \label{fig:AUC_freq}
\end{figure} 

\subsection{Choice of the frequency band}
\label{res:freq}
As seen in Figure \ref{fig:AUC_freq}, the frequency bands that provided the highest discriminant rates in terms of AUC were 8-24 Hz and 8-30 Hz. Since the latter is more susceptible any muscular artifacts~\cite{vanDeVelde1998}, we preferred to set the bandpass cut-off frequencies between 8-24 Hz.
Low frequencies ($<$4 Hz) were discarded not only for their moderate discrimination power, but also because in a real implementation of the BVI this spectral range would be corrupted by electromagnetic interferences inherent in clinical environments. Other frequency combinations were not superior to 8-24 Hz classification rates.

\subsection{Comparison of distance metrics}
The first test used the 14 most central electrodes, thus covariance matrices were obtained from EEG signals $\mathbf{X}  \in \mathbb{R}^{1250 \times 14}$. The classifier settings were $L=25$ and $K=3$. 

\begin{figure}[t!] 
   \centering
   \includegraphics[trim = 0cm 0.5cm 1cm 0.75cm,clip=true,width=\linewidth]{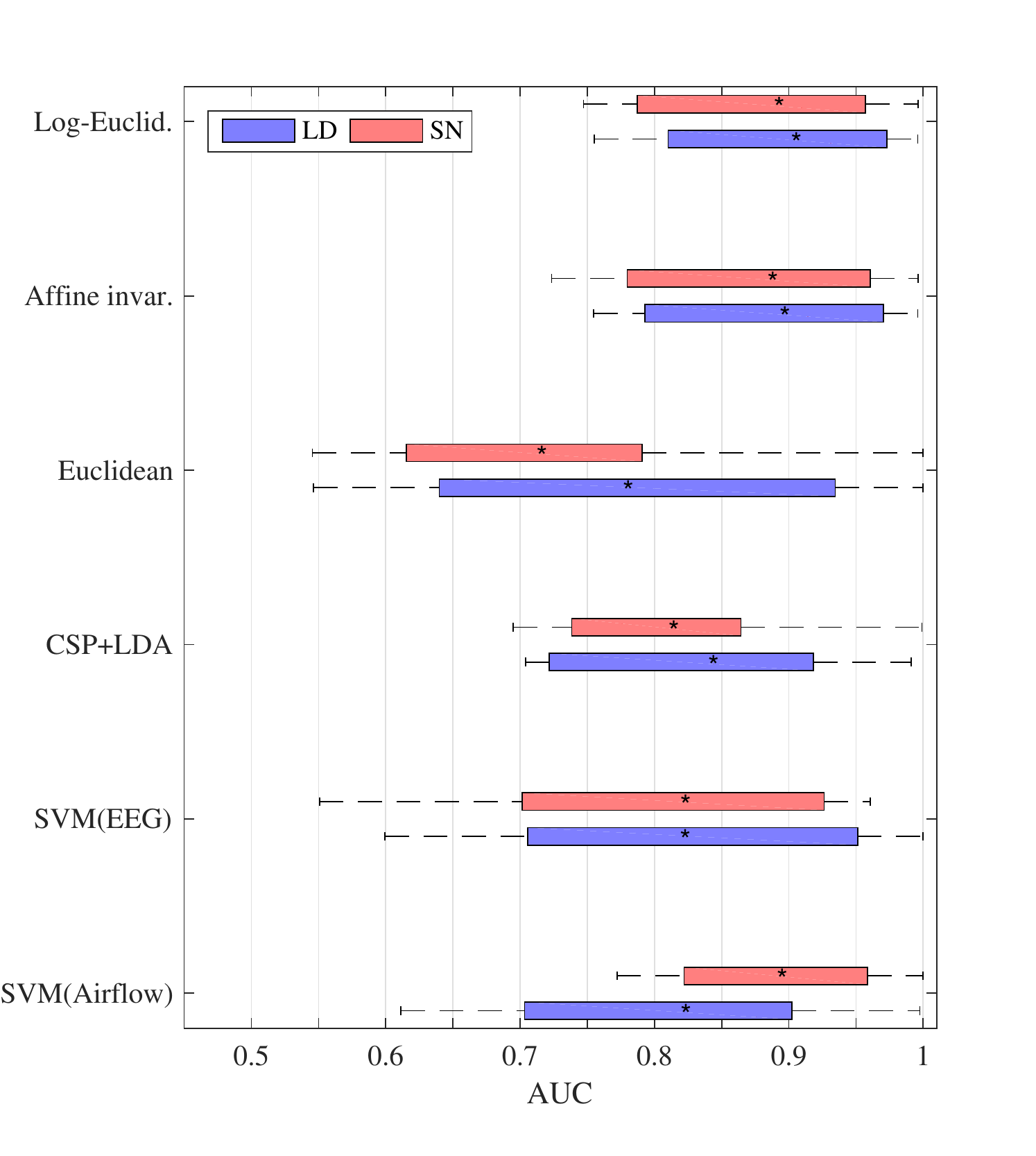} 
   \caption{BVI performances for detecting LD (blue) and SN (red) conditions. Boxplots display the areas under ROC curves (AUCs) for different detectors using the EEG and air flow signals (with $L=25$ and $K=3$). Asterisks (*) denote the mean AUC values across subjects.}
   \label{fig:AUC_ini}
\end{figure}

Figure \ref{fig:AUC_ini}  shows that the best performing metric is the Log-Euclidean, which provided slightly better AUC values than Riemannian distances in LD condition ($\text{AUC}=0.91$ versus $\text{AUC}=0.90$) and almost equal AUC values ($\text{AUC}\approx0.9$) for the SN condition. The classification performances for SN had the largest variability probably due to the discontinuous occurrence of sniffs (one every two breaths). Although the performances of Log-Euclidean are similar to those obtained by the affine invariant  Riemannian metric, the former is advantageous from the computational point of view due to its reduced algorithmic complexity  (CPU times were\footnote{Using a 2.8GHz dual-core Intel Core i7 processor, 16GB of memory}, on average, 3 time faster).   

Euclidean distances provided the lowest average classification rates and the most unstable AUC values, demonstrating the limitation of linear matrix operators (i.e. the arithmetic mean) for EEG covariance matrix classification. 

\begin{figure*}[t!] 
   \centering
   \resizebox{0.98\linewidth}{!}{\includegraphics[trim = 4cm 0cm 3.8cm 0.4cm,clip=true,width=\linewidth]{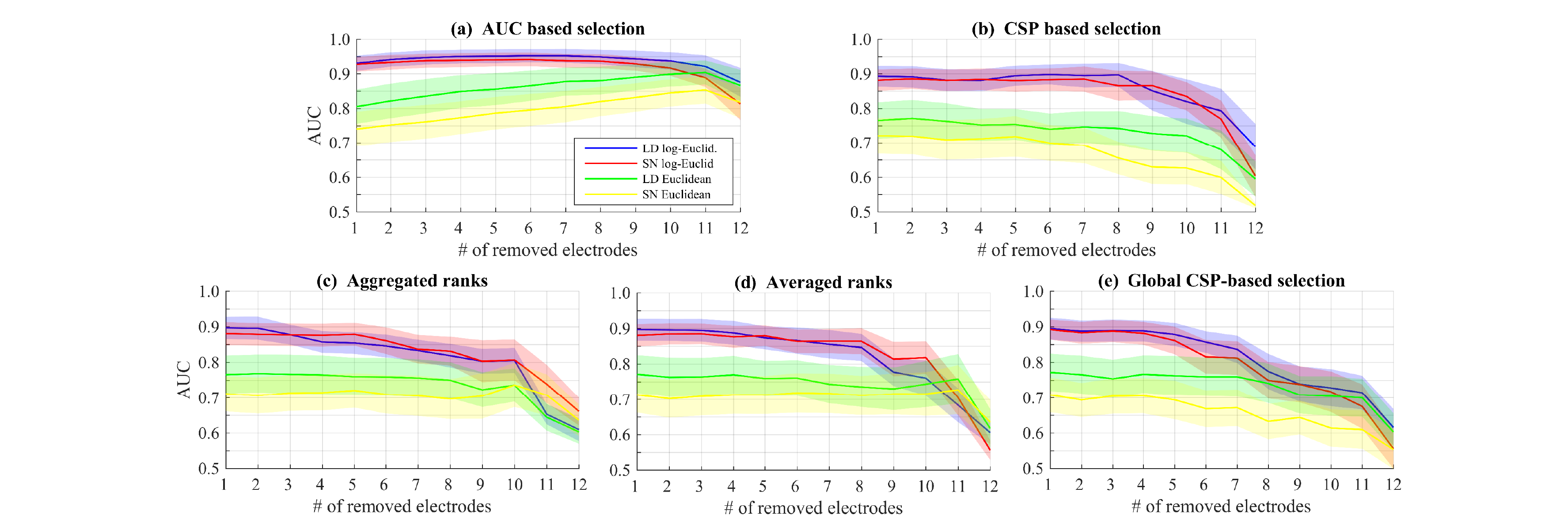}} 
    \caption{Classification performance when individual optimization (a and b) or a common electrode set-up (c-e) are applied. AUC values are expressed as mean (solid lines) $\pm$ standard error (shadowed areas). }
    \label{fig:AUC_elect}
    \end{figure*}

\subsection{Comparison with standard BCI methods}
The analysis of $\text{AUC}$ values also proves that Riemannian geometry is a better framework to classify EEG covariance matrices, than the outlier detector based on one-class SVMs. We notice, however, that the latter provided better results than the classifier with Euclidean structure (mean values are $\text{AUC}= 0.82$ versus $\text{AUC}\leqslant0.77$). 

In LD condition, classification performances obtained with air flow ($\text{AUC}=0.81$) remained below than those obtained using log-Euclidean and affine invariant distances, and almost equal to the one-class SVMs EEG-based detector ($\text{AUC}=0.82$). Because of the characteristic large pattern of air flow during sniffs, air flow signals provided better classification in SN than SVM EEG-based detector ($\text{AUC}=0.89$ and $\text{AUC}=0.81$, respectively) and similar performances to our approach.

Finally, if we considered the BVI as a two-class detector (unsuitable for on-line detection as discomfort classes are not known a priori), the classification by CSP and  LDA yielded $\text{AUC}=0.84$ for LD and $\text{AUC}=0.81$ for SN, values clearly below our Riemannian geometry based approach.

\subsection{Choice of electrodes}
\label{elect_rankings}
We assessed the most relevant electrodes in the proposed BVI, starting from a set of 14 central electrodes from the original 32-electrode montage. Retained electrodes included positions on primary sensorimotor cortices (C3, C4),  higher-level secondary and association cortices (Fz and Cz), pre-parietal (CP1 and CP2) and pre-frontal cortices (F3, F4), pre-motor and supplementary motor areas (FC1, FC2), central (FC5 and FC6) and parieto-occipital areas (CP5 and CP6). The choice of these areas follows results of earlier experiments describing the cortical networks elicited during respiratory load compensation~\cite{macefield1991, raux2007, leupoldt2010, raux2013, morawiec2015}. 

According to the CHOrRA procedure to find the best channels from the initial set-up (c.f Algorithm 1 in Section \ref{ch_selelct}), we first found intra-subject rankings and then applied a global rank aggregation. As described above, we employed AUC as the measure of classification performance and performed the rankings from $V$=10 lists (i.e. 10 different learning periods) of each subject.

\subsubsection{Intra-subject optimal selection}
The classification results are shown in Figure~\ref{fig:AUC_elect}-(a) for intra-subject AUC optimized ranks and Figure~\ref{fig:AUC_elect}-(b) for intra-subject CSP electrode ranks, where AUC values are expressed as a function of the number of removed electrodes. In both cases, the evolution of the curves during the optimization process display a slight increasing tendency of AUC for the Log-Euclidean metric when electrodes contributing negatively to classifications are removed. Classification performances decrease with configurations smaller than six electrodes. 

Results show that customized individual selection of electrodes can provide an optimal AUC within a single subject. Using an intra-subject AUC based optimization, classification rates can be improved up to 0.95 using the six most significant electrodes. On the other hand, selecting the electrodes according to CSP patterns provides lower AUCs, with the best 6-electrode reaching an average AUC value of 0.85 in LD.    

As shown in different plots of Figure~\ref{fig:AUC_elect}, for a configuration with more than three electrodes, the Log-Euclidean metric performs better than the Euclidean distance at every optimization step. These results support the idea that large spatio-temporal information of the EEG (as reflected by the sample covariance matrices) is optimally captured when the intrinsic geometry structure of the underlying data is taken into account.

\subsubsection{Inter-subject optimal selection}
We applied the two rank combinations proposed by CHOrRa to the lists of sorted electrodes of all subjects to find a general ranking of electrodes for each method.  Following the rank of the final lists, we computed the AUC values by reducing the initial 14-electrode set to two sensors (see Figure \ref{fig:AUC_elect}-(c-d)). Classifications provided by inter-subject CSP ranks are depicted in Figure \ref{fig:AUC_elect}-(e).
 
Results from both CHOrRA procedures indicate that a good compromise between a reduced number of electrodes and reasonable classifications can be reached by selecting the best 6 electrodes obtained by the averaged ranks ($\text{AUC} \geqslant 0.85$ for both LD and SN conditions). In line with intra-subject ranks, the general electrode list obtained by CSP patterns resulted in poorer classification rates, providing $\text{AUC} \leqslant 0.80$ for a set of 6 electrodes.

Results indicate that, for general configurations with more than three electrodes, Log-Euclidean metric still performs better than the Euclidean distance.

\subsubsection{Neurophysiological Interpretation}
\begin{figure}[t!] 
   \centering
   \includegraphics[trim = 2cm 2cm 2cm 0.5cm,clip=true,width=0.98\linewidth]{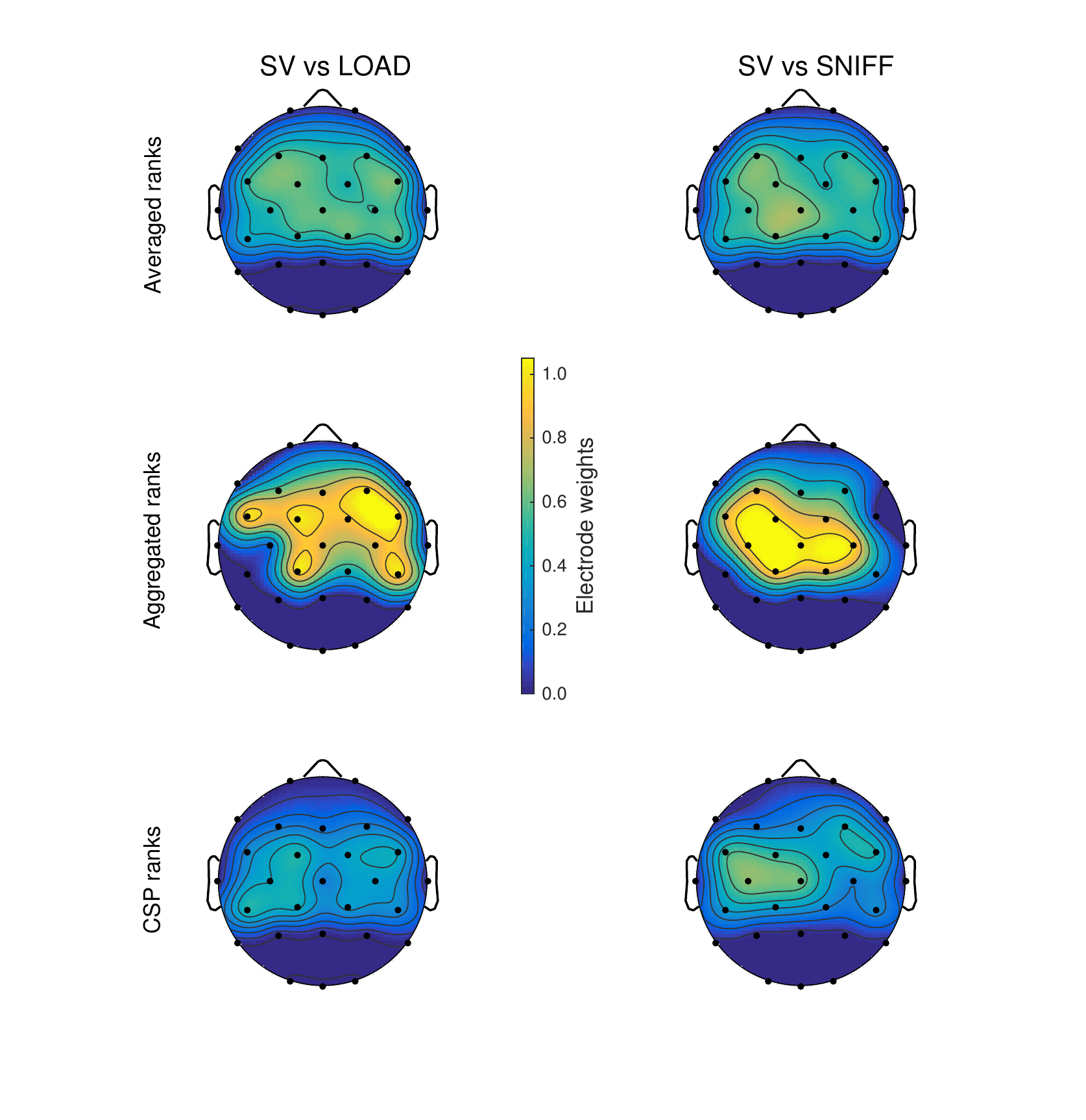} 
\caption{Topographic plots of influence scores of channels in the classifications of LD and SN conditions. Upper and middle plots are obtained through the CHOrRA method (with Log-Euclidean metric, $L=25$ and $K=3$) whereas bottom plots correspond to the averaged, normalized weights from CSP patterns. }
    \label{fig:heads}
    \end{figure}

The aggregated CHOrRa scores displayed in Figure~\ref{fig:heads} indicate scalp zones with more influence for the classification of both SN and LD conditions. Although both optimization procedures (robust rank aggregation and rank averaging) provide similar classification performance, the spatial concentration for discrimination of scalp regions obtained by the robust aggregation method is larger than those obtained by a classical averaging rank procedure. 

The spatial distribution of scores and the rankings (see Figure \ref{fig:heads}) indicates that discrimination of  LD condition is better if the electrode configuration contains the pre-motor and supplementary motor area (FC1), the fronto-central region (FC6) and the supramarginal gyrus (the part covered by CP6 electrode). This agrees with previous findings suggesting that during LD, the supplementary motor area (SMA) is recruited~\cite{raux2007, raux2007b}, most likely within the frame of a cortico-subcortical cooperation allowing compensation of the inspiratory load~\cite{georges2016}. 

During SN, the conscious preparation of a breath activates pre-motor areas and the execution of the breath activates motor areas~\cite{macefield1991, raux2007}. Indeed, our results show that SN condition is better discriminated if electrodes include the supplementary motor area (FC1), the central motor area Cz and part of the somatosensory association cortex (covered by the electrode CP1). 

Figure \ref{fig:heads} also shows the relative weights of electrode positions related to the most relevant CSP patterns.  In both LN and SN conditions, topographic plots reveal that highly ranked positions reside in similar scalp regions. For loaded breathing, CSP weights suggest a bilateral activation of the cortex that includes fronto-central, central (motor cortex) and some centro-parietal positions (somato-sensory cortex). On the other hand, during voluntary sniffs, highly scored electrodes match fronto-central positions on the left hemisphere (motor and pre-motor areas). 

Electrode selection based on CSP coefficients converges to a great extent with CHOrRA selection. Nevertheless, for a given number of electrodes our proposed approach provides better classification performances than the counterpart selection in CSP ranks (see Figure \ref{fig:AUC_elect}).

To confirm that changes in brain activity during LD and SN epochs detected by the EEG-based classifier are related to breathing, correlations between EEG and air flow were also computed after correcting by autocorrelations and time trends present in time series~\cite{Jenkins1968}.  Results show that mean correlations in central areas (including C3, C4, Fz and FC1 electrodes) increase about $20\%$. 
The increase of correlation values between EEG and air flow signals confirms that changes in brain state detected by our classifier are related to respiration. Our results suggest that, in general,  EEG signals might provide better discriminant features than air flow to detect breathing discomfort.  


\begin{figure}[b!] 
   \centering
  \resizebox{\linewidth}{!}{\includegraphics[trim = 2.3cm 1cm 1cm 2cm,clip=true,width=\linewidth]{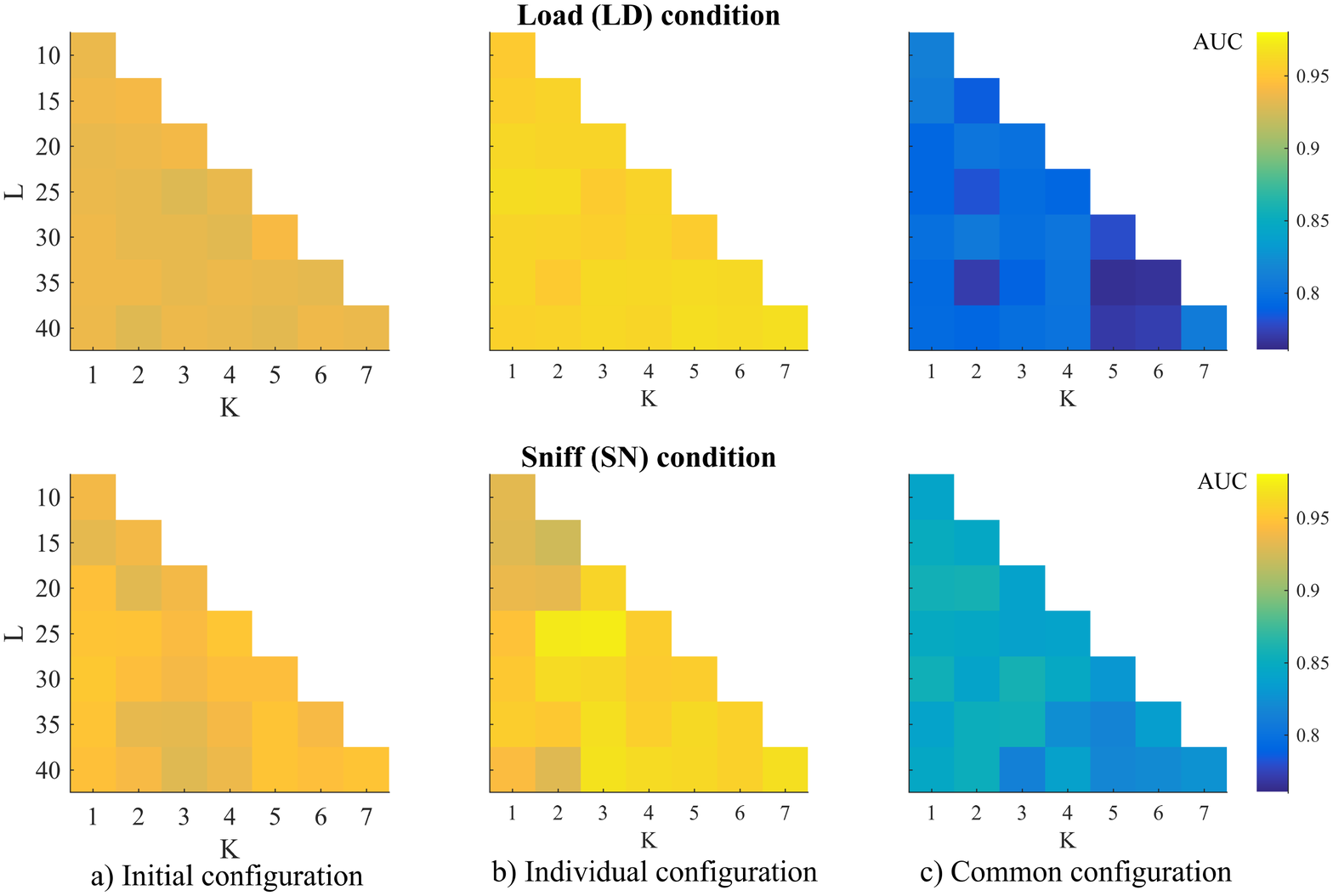}}
      \caption{AUC values (colors in boxes) against the number of prototypes ($K$ centroids) and the learning time ($L$ CMs) obtained for classification of the LD and SN conditions with the Log-Euclidean metrics. The ladder-shaped grids are explained by the limitation of the clustering algorithm to compute reliable centroids when few CMs are available.}
    \label{fig:AUC_KL}
\end{figure}
              
\subsection{Training time vs. number of prototypes}
Once the effect of different metrics and channels configurations were assessed, we tested the influence of the learning parameters, i.e. the number of covariance matrices used as a reference ($L$) and the number of prototypes ($K$) used to characterize the reference class.

For this, we assessed classification performance for different values of $K$, ranging from 1 to 7 and  and for different number of learning matrices, $L$, from 20 to 40 in steps of 5. This procedure was also repeated with 10-fold cross validation for LD and SN conditions. Three different electrode configurations were tested: 1) the initial 14-electrode configuration, 2) a selection of 6 electrodes optimized for every subject and 3) a general configuration with the best 6 electrodes after applying the global channel selection procedure. 

Results are depicted in the color maps of Figure \ref{fig:AUC_KL}, where average AUC values are expressed as a function of $K$ and $L$. Importantly, this figure shows a small effect of both parameters on overall classification rates, regardless of the electrode configuration. In view of these findings, the choice of a small number of prototypes ($K \le 3$) and a short learning time ($L=20$ CMs) is an advantageous trade-off between computational complexity and classification performances, even for a common electrode configuration setting.

\subsection{Strengths and limitations}
Compared to classification based on air flow, our findings demonstrate a better discriminatory power of the covariance patterns of EEG signals to detect a patient-ventilator disharmony ($\text{AUC} = 0.91$ versus $\text{AUC} = 0.81$, in average, for the loaded breathing condition).  The analysis of AUC values suggests that Riemannian geometry is a better framework to manipulate covariance matrices, even with a few number of channels, than classical matrix operators with an Euclidean structure. We notice that one-class SVMs detector also provides better results than a classifier employing Euclidean metrics. 

In contrast with other classical methods used in BCI, (as CSP which assumes a-priori labelled data from two classes), the results support the better discriminant capacity of our approach to identify anomalous respiratory periods, by learning from a single training set containing only normal respiratory epochs.

The proposed recursive channel selection procedure may provide a subject-customized BVI setting with a reduced number of six channels and maintained performance ($\text{AUC} \approx 0.95$). If the channel configuration includes the most significant electrodes across subjects, classification rates are reduced compared to customized optimization. Nevertheless, this general setting provides a good compromise between a reduced number of electrodes (6 channels) and reasonable classification performances ($\text{AUC}\geqslant 0.85$), which is advantageous in general clinical practice where ready-to-use devices are necessary. Interestingly, overall classification performance do not significantly depend on the other parameters of the classifier (learning time and number of prototypes). 
Similar to our approach, there have been previous attempts to use divergences from information  theory (e.g. KL divergence, Jensen-Shannon divergences or Bregman matrix divergence) for developing distances on covariance matrices \cite{samek2014}. The use of such dissimilarity measures could improve the performance of our algorithm.
An optimization of control parameters for other SVMs models might also increase classification performances but this subject is beyond the scope of our paper.

One major limitation of our approach results from the difficulty to properly estimate the covariance matrices. Poor estimates do not accurately reflect the underlying neural process and thus will directly affect the classifier's performances. The estimation of a covariance matrix in an EEG segment assumes a multivariate, stationary
stochastic process so it can be strongly affected by artifacts such as eye blinks, muscle activity or swallowing. Changes over longer time scales such as changes in electrode impedance, lose of electrodes or sudden electrode shifts, would also deteriorate classification performances. The use of regularization or adaptive learning approaches (see \cite{lotte2015} and references therein) could improve the robustness of the algorithm.
\vspace{0.5cm}

\section{Conclusion}
\label{concl}
This work provides the first extensive evaluation of what could be a brain-ventilator interface (BVI) designed to detect altered respiratory conditions in patients on mechanical ventilation. The novelty of our proposal is a Riemannian Geometry approach to identify a cortical signature of breathing discomfort from  spatio-temporal EEG patterns.

In general, characterization of brain networks provides meaningful insights into the functional organization of cortical activities underlying breathing control and respiratory diseases. Our study supports the hypothesis of a strong correlation between voluntary and compensatory respiratory efforts and the activation of cortical circuits~\cite{macefield1991, raux2007, leupoldt2010}. Correlation increase between EEG and air flow during breathing discomfort epochs also corroborates that changes in brain state detected by our classifier are related to respiration. A hybrid BCI combining EEG and airflow signals could be explored in future studies.

The introduction of a BVI is a first step toward a critical class of interfaces for respiratory control applications in a variety of clinical conditions where the use of mechanical ventilation is required to decrease the work of breathing in the patients. For instance, in patients suffering amyotrophic lateral sclerosis, we have recently reported a respiratory-related cortical activity~\cite{georges2016}. Effective translation of our approach to a suitable device for long-term monitoring of these patients faces difficult challenges that arise from the nature of the heterogeneous population (e.g. different respiratory problems) and the numerous sources of artifacts in clinical units. Nevertheless, due to its technical simplicity (portable, non-invasive, with few electrodes and fast computation) the proposed BVI can be highly operable in clinical environments as well as in custom-designed systems. Indeed, our algorithm~\cite{covemPatent} is currently assessed by a large clinical study undertaken at the ICU of our group. Beyond short-term applications where the BVI would prompt clinicians to run a dyspnea check-list, future work should also integrate the BVI in a feedback scheme to automatically set ventilator parameters with minimal physician intervention.

\section*{Acknowledgments}
This work was supported by the program Investissement d'Avenir ANR-11-EMMA-0030 and ANR-10-AIHU 06 of the French Government and by the grant Legs Poix from the Chancellerie de l'Universit\'e de Paris, France. X. Navarro-Sune is financially supported by Air Liquide Medical Systems S.A., France. A.L. Hudson was supported by an NHMRC (Australia) Early Career Fellowship.
\bibliographystyle{ieeetr}

\begin{thebibliography}{10}
\footnotesize
\bibitem{carlucci2001}
A.~Carlucci et al. {SRLF
  Collaborative Group on Mechanical Ventilation}, ``Noninvasive versus
  conventional mechanical ventilation. an epidemiologic survey,'' {\em American
  Journal of Respiratory and Critical Care Medicine}, vol.~163, no.~4,
  pp.~874--880, 2001.

\bibitem{leung1997}
P.~Leung et al. ``Comparison of assisted ventilator modes
  on triggering, patient effort, and dyspnea,'' {\em American Journal of
  Respiratory and Critical Care Medicine}, vol.~155, no.~6, pp.~1940--1948,
  1997.

\bibitem{Sinderby1999}
C.~Sinderby et al. ``Neural control of mechanical ventilation in
  respiratory failure,'' {\em Nature Medicine}, vol.~5, no.~12, pp.~1433--1436,
  1999.

\bibitem{Spahija2010}
J.~Spahija et al. , ``Patient-ventilator interaction during pressure support
  ventilation and neurally adjusted ventilatory assist,'' {\em Critical care
  medicine}, vol.~38, no.~2, pp.~518--526, 2010.

\bibitem{raux2007}
M.~Raux et al. ``Electroencephalographic evidence for pre-motor cortex
  activation during inspiratory loading in humans,'' {\em The Journal of
  Physiology}, vol.~578, no.~Pt 2, pp.~569--578, 2007.

\bibitem{raux2007b}
M.~Raux et al.
  ``Cerebral cortex activation during experimentally induced ventilator
  fighting in normal humans receiving noninvasive mechanical ventilation,''
  {\em The Journal of the American Society of Anesthesiologists}, vol.~107,
  no.~5, pp.~746--755, 2007.

\bibitem{Grave2013}
R.~Grave~de Peralta et al. ``Patient machine
  interface for the control of mechanical ventilation devices,'' {\em Brain
  sciences}, vol.~3, no.~4, pp.~1554--1568, 2013.

\bibitem{Varela2001}
F.~Varela et al ``The brainweb:
  phase synchronization and large-scale integration,'' {\em Nature Reviews
  Neuroscience}, vol.~2, no.~4, pp.~229--239, 2001.

\bibitem{barachant2012}
A.~Barachant et al. ``Multiclass brain-computer
  interface classification by riemannian geometry,'' {\em {IEEE} Transactions
  on Biomedical Engineering}, vol.~59, no.~4, pp.~920--928, 2012.

\bibitem{mak2009}
J.~N. Mak and J.~R. Wolpaw, ``Clinical applications of brain-computer
  interfaces: current state and future prospects,'' {\em IEEE Reviews in Biomedical
  Engineering}, vol.~2, pp.~187--199, 2009.

\bibitem{hudson2016}
A.~L. Hudson et al., ``Electroencephalographic detection of respiratory-related
  cortical activity in humans: from event-related approaches to continuous
  connectivity evaluation,'' {\em Journal of Neurophysiology},   vol.~115,  pp.~2214--2223,
  2016.

\bibitem{dodek2004}
P.~Dodek et al., ``Evidence-based clinical practice guideline for the prevention 
  of ventilator-associated pneumonia,'' {\em Annals of Internal Medicine},   vol.~141, no.~4,   pp.~305--313, 2004.

\bibitem{pfurtscheller1999}
G.~Pfurtscheller and F.~L. Da~Silva, ``Event-related eeg/meg synchronization
  and desynchronization: basic principles,'' {\em Clinical neurophysiology},
  vol.~110, no.~11, pp.~1842--1857, 1999.

\bibitem{moakher2005}
M.~Moakher, ``A differential geometric approach to the geometric mean of
  symmetric positive-definite matrices,'' {\em {SIAM} Journal on Matrix
  Analysis and Applications}, vol.~26, no.~3, pp.~735--747, 2005.

\bibitem{bergerBOOK}
M.~Berger, {\em A panoramic view of Riemannian geometry}.
\newblock Springer Science \& Business Media, 2012.

\bibitem{pennec2006}
X.~Pennec et al., ``A {R}iemannian framework for tensor
  computing,'' {\em International Journal of Computer Vision}, vol.~66, no.~1,
  pp.~41--66, 2006.

\bibitem{arsigny2006}
V.~Arsigny et al., ``Log-{E}uclidean metrics for
  fast and simple calculus on diffusion tensors,'' {\em Magnetic Resonance in
  Medicine}, vol.~56, no.~2, pp.~411--421, 2006.

\bibitem{dryden2009}
I.~L. Dryden et al.,``Non-{E}uclidean statistics for
  covariance matrices, with applications to diffusion tensor imaging,'' {\em
  The Annals of Applied Statistics}, pp.~1102--1123, 2009.

\bibitem{vemuri2011}
B.~C. Vemuriet al., ``Total {B}regman divergence
  and its applications to dti analysis,'' {\em  IEEE Transactions on Medical Imaging}, vol.~30, no.~2, pp.~475--483, 2011.

\bibitem{cherian2013}
A.~Cherian et al., ``Jensen-{B}regman
  logdet divergence with application to efficient similarity search for
  covariance matrices,'' {\em IEEE Transactions on Pattern Analysis and Machine Intelligence}, vol.~35, no.~9, pp.~2161--2174, 2013.

\bibitem{arsigny2007}
V.~Arsigny et al., ``Geometric means in a novel
  vector space structure on symmetric positive-definite matrices,'' {\em SIAM
  Journal on Matrix Analysis and Applications}, vol.~29, no.~1, pp.~328--347,
  2007.

\bibitem{karcher1977}
H.~Karcher, ``Riemannian center of mass and mollifier smoothing,'' {\em
  Communications on Pure and Applied Mathematics}, vol.~30, no.~5,
  pp.~509--541, 1977.

\bibitem{fletcher2004}
P.~T. Fletcher and S.~Joshi, ``Principal geodesic analysis on symmetric spaces:
  Statistics of diffusion tensors,'' in {\em Computer Vision and Mathematical
  Methods in Medical and Biomedical Image Analysis}, no.~3117, pp.~87--98,
  2004.

\bibitem{lotte2007}
F.~Lotte et al.,  ``A review
  of classification algorithms for {EEG}-based brain--computer interfaces,'' {\em
  Journal of neural engineering}, vol.~4, no.~2, p.~R1, 2007.

\bibitem{scholkopfBook}
A.~J. Smola and B.~Sch{\"o}lkopf, {\em Learning with kernels}.
\newblock The MIT Press Cambridge, MA, USA, 2001.

\bibitem{quang2013}
M.~H. Quang et al.,  ``A unifying framework for
  vector-valued manifold regularization and multi-view learning,'' in {\em
  Proceedings of the 30th International Conference on Machine Learning
  (ICML-13)}, pp.~100--108, 2013.

\bibitem{abaza2009}
A.~A. Abaza et al., ``Classification of voluntary
  cough sound and airflow patterns for detecting abnormal pulmonary function,''
  {\em Cough}, vol.~5, no.~1, p.~1, 2009.

\bibitem{blankertz2008}
B.~Blankertz  et al., ``Optimizing
  spatial filters for robust {EEG} single-trial analysis,'' {\em IEEE Signal
  Processing Magazine}, vol.~25, no.~1, pp.~41--56, 2008.

\bibitem{haufe2014}
S.~Haufe  et al.,  ``On the 
interpretation of weight vectors of linear models in multivariate neuroimaging,'' 
{\em Neuroimage}, vol.~87, no.~1, pp.~96--110, 2014.

\bibitem{voipio2003}
J.~Voipio et al., 
  ``Millivolt-scale dc shifts in the human scalp {EEG}: evidence for a
  nonneuronal generator,'' {\em Journal of neurophysiology}, vol.~89, no.~4,
  pp.~2208--2214, 2003.

\bibitem{ramoser2000}
H.~Ramoser et al., ``Optimal spatial
  filtering of single trial {EEG} during imagined hand movement,'' {\em
  IEEE Transactions on Rehabilitation Engineering}, vol.~8, no.~4,
  pp.~441--446, 2000.

\bibitem{farquhar2006}
J.~Farquhar et al., ``Regularised {CSP} for sensor selection in {BCI},'' in {\em
 Proceedings of the 3rd International Brain-Computer Interface. Workshop and Training Course 2006. Graz University of Technology, Austria}, pp.~14--15, 2006.

\bibitem{wang2005}
Y.~Wang et al.,  ``Common spatial pattern method for channel
  selection in motor imagery based brain-computer interface,'' in {\em IEEE
  Engineering in Medicine and Biology 27th Annual Conference}, 2005.

\bibitem{lal2004}
T.~N. Lal et al.,  ``Support vector channel selection in
  {BCI},'' {\em IEEE Transactions on Biomedical Engineering}, vol.~51, no.~6,
  pp.~1003--1010, 2004.

\bibitem{kolde2012}
R.~Kolde et al.,  ``Robust rank aggregation for gene
  list integration and meta-analysis,'' {\em Bioinformatics}, vol.~28, no.~4,
  pp.~573--580, 2012.

\bibitem{ray2015}
A.~M. Ray et al.,  ``A subject-independent pattern-based brain-computer interface,'' {\em Frontiers
  in behavioral neuroscience}, vol.~9, 2015.

\bibitem{vanDeVelde1998}
M.~van~de Velde et al.,  ``Detection of muscle
  artefact in the normal human awake {EEG},'' {\em Electroencephalography and
  Clinical Neurophysiology}, vol.~107, no.~2, pp.~149--158, 1998.

\bibitem{Jenkins1968} 
G.~M.~Jenkins and D.~G.~Watts, 
{\em Spectral analysis and its applications}.
\newblock  Holden-Day, San Francisco, California, 1968.

\bibitem{georges2016}
M.~Georges  et al., ``Cortical drive to breathe in amyotrophic lateral sclerosis: 
a dyspnoea-worsening defence?,'' 
{\em European Respiratory Journal}, ERJ--01686, 2016.

\bibitem{macefield1991}
G.~Macefield and S.~C. Gandevia, ``The cortical drive to human respiratory
  muscles in the awake state assessed by premotor cerebral potentials,'' {\em
  The Journal of Physiology}, vol.~439, pp.~545--558, 1991.

\bibitem{leupoldt2010}
A.~von Leupoldt et al., ``Cortical sources of the respiratory-related evoked potential,''
  {\em Respiratory physiology \& neurobiology}, vol.~170, no.~2, pp.~198--201,
  2010.

\bibitem{raux2013}
M.~Raux  et~al.,
  ``Functional magnetic resonance imaging suggests automatization of the
  cortical response to inspiratory threshold loading in humans,'' {\em
  Respiratory physiology \& neurobiology}, vol.~189, no.~3, pp.~571--580, 2013.

\bibitem{morawiec2015}
E.~Morawiec et al.,
  ``Expiratory load compensation is associated with electroencephalographic
  premotor potentials in humans,'' {\em Journal of Applied Physiology},
  vol.~118, no.~8, pp.~1023--1030, 2015.

\bibitem{samek2014}
W.~Samek et al., ``Divergence-based framework for common spatial patterns algorithm,'' 
{\em IEEE Reviews in Biomedical Engineering}, vol.~7, pp~50--72, 2014.

\bibitem{lotte2015}
F.~Lotte, ``Signal processing approaches to minimize or suppress calibration time in oscillatory activity-based brain computer interfaces'',
{\em Proceedings of the IEEE}, vol~103, no~6, pp~871--890, 2015.



\bibitem{covemPatent}
T.~Similowski et al.,``Method for
  characterising the physiological state of a patient from the analysis of the
  cerebral electrical activity of said patient, and monitoring device applying
  said method,'' 2013.
\newblock WO Patent WO2013164462 (A1).

\end{thebibliography}

\end{document}